\documentclass[aps,prl,twocolumn,groupedaddress]{revtex4}
\usepackage{graphicx}
\usepackage{epstopdf}
\begin{document}

\preprint{quant-ph/0507076}
\title{Measuring non-linear functionals of quantum harmonic oscillator states.}

\author{K. L. Pregnell}
\email{pregnell@physics.uq.edu.au}
\thanks{We thank D. T. Pegg and J. Dodd for useful discussions relating to the manuscript and A. Lund for proof reading the manuscript.  This work was supported by the Australian Research Council.}
\affiliation{Centre for Quantum Computer Technology, Department of Physics\\
University of Queensland, Brisbane, QLD 4072, Australia}
\date{\today}

\begin{abstract}
Using only linear interactions and a local parity measurement we show how entanglement can be detected between two harmonic oscillators. The scheme generalizes to measure both linear and non-linear functionals of an arbitrary oscillator state.  This leads to many applications including purity tests, eigenvalue estimation, entropy and distance measures - all without the need for non-linear interactions or complete state reconstruction.  Remarkably, experimental realization of the proposed scheme is already within the reach of current technology with linear optics.
\end{abstract}

\pacs{03.67.-a, 03.65.Ud, 42.50.-p, 42.50.Dv}
\keywords{Entanglement witness, non-linear functionals, swap operation, harmonic oscillator, linear optics}
\maketitle

In the context of quantum communication and computing protocols, measures such as purity, fidelity and entanglement characterize the performance and non-classical resources in a physical experiment.  Such measures not only provide a link with theoretical models, but also provide standards for defining benchmarks \cite{Caves:2004}. With continuing technical developments in these areas, it is becoming increasingly necessary to identify practical and efficient schemes to measure such quantities. 

One obvious method is to first reconstruct the complete density matrix from a series of measurements using, for example, the well-know procedure of quantum state tomography \cite{Smithey:1993,Leibfried:1996,Lvovsky:2001}. From this reconstructed density matrix the desired measure can then be computed. Although this technique is realizable, it is not efficient as much more information about the quantum state is obtained than is actually needed.  

A more direct method was recently proposed by Filip \cite{Filip:2002}, and expanded upon by others \cite{Ekert:2002,Horodecki:2003,Carteret:2005}.  They proposed a specific quantum circuit capable of measuring the non-linear functional $\mathrm{Tr}(\rho^{k}_{a}\rho^{l}_{b})$ of the density matrices $\rho_{a}$ and $\rho_{b}$.  The scheme requires as inputs, $k$ and $l$ copies of the states $\rho_{a}$ and $\rho_{b}$, respectively. It was independently shown by Brun \cite{Brun:2004} that any polynomial function of a state up to degree $q$ can be estimated by a joint measurement on $q$ copies of the system.  Amazingly, when the systems are entangled the measurement doubles as an entanglement witness \cite{Horodecki:2002}, giving a negative value for some entangled states while ensuring a positive value for all separable states. 

The circuit, although elegant, unfortunately requires a nontrivial interaction between a control qubit and the $k$ and $l$ copies of each system - the targets. In its simplest form with two targets, $k+l=2$, the interaction must generate a controlled-\textsc{swap} operation between the control and the two targets.  Such an operation does nothing to the targets if the control is in the logical zero state and applies the unitary swap operation
\begin{equation}\label{1}
      V|\phi_{1}\rangle|\phi_{2}\rangle=|\phi_{2}\rangle|\phi_{1}\rangle
\end{equation}
to the two targets if the control is in the logical one state, where here $|\phi_{1}\rangle$ and $|\phi_{2}\rangle$ are arbitrary target states. Implementing this gate is a challenging requirement for any experimental architecture, especially when the target states are of arbitrary dimension. In the case of a harmonic oscillator where the dimension of the Hilbert space is infinite, implementing this gate using, for example, non-deterministic KLM-type gates \cite{KLM, OBrien:2003} is only possible when there is at most one quanta of energy in each oscillator. 

In this paper we propose an alternative to the Filip scheme for harmonic oscillators which, remarkably, requires only \emph{linear} coupling between different oscillators making it realizable with current technology.  Like the Filip proposal, the scheme can measure both linear and non-linear functionals of the density matrix, while for entangled oscillators the scheme acts as an entanglement witness. We begin by first reviewing some of the properties of the swap operator \cite{Ekert:2002,Horodecki:2003,Brun:2004,Carteret:2005}, along with its generalization to multi-particles, and show how it relates to a host of relevant information measures.  

For the swap operator defined in Eqn.~(\ref{1}) it is straightforward to show that for two separable states $\rho_{a}$ and $\rho_{b}$, 
\begin{equation}\label{2}
    \mathrm{Tr}(\rho_{a}\rho_{b})=\mathrm{Tr}_{ab}(\rho_{a}\otimes\rho_{b}  V).
\end{equation}
Although this seems like a trivial relation, it suggests that a direct way to measure the overlap between two unknown states is to measure the expectation value of the swap operator $\langle   V\rangle$. In the case where both systems are in the same state $\rho_{i}$, the expectation value is equivalent to the purity of the system, $\langle   V\rangle=\mathrm{Tr}(\rho^{2}_{i})$. From the purity and the overlap one could obtain, for example, the Hilbert-Schmidt distance $\mathrm{Tr}[(\rho_{a}-\rho_{b})^{2} ]$ .

In the case where the state is not separable Eqn.~(\ref{2}) is no longer valid. To illustrate this take, for example, the entangled state $(|\alpha\rangle|\beta\rangle-|\beta\rangle|\alpha\rangle)/\sqrt{2}$, where $|\alpha\rangle$ and $| \beta\rangle$ are orthonormal.  From (\ref{1}) the expectation value of the swap operator is $-1$ which, in contrast to (\ref{2}), is negative. This is a example of the separability criteria of Horodecki, Horodecki and Horodecki \cite{Horodecki:1996} which states that a density matrix $\rho$ is entangled iff there exists a Hermitian operator $H$, an entanglement witness, such that 
\begin{equation}\label{3}
    \mathrm{Tr}(\rho   H)<0,
\end{equation}
while for all separable states $\rho_{sep}$
\begin{equation}\label{4}
    \mathrm{Tr}(\rho_{sep}  H)\ge 0.
\end{equation}
From Eqn.~(\ref{2}) and by explicit example we see that the swap operator is an example of an entanglement witness. A measurement of $\langle V\rangle$ is said to have witnessed the entanglement when the outcome $\langle V\rangle <0$. In fact, it can be shown that the swap operator is an \emph{optimal} entanglement witness in the sense that it forms a hyper-plane that is tangential to the convex set of separable states \cite{Bertlmann02,Bertlmann05}. That is a separable state $\rho_{sep}'$ exist such that $\mathrm{Tr}[ \rho_{sep}' V]=0$.  Such a state is $|\alpha\rangle\otimes|\beta\rangle$.

A generalization of the swap operator to multiple systems is defined as
\begin{equation}\label{5}
      V_{N}|\phi_{1}\rangle|\phi_{2}\rangle\dots|\phi_{N}\rangle=|\phi_{N}\rangle|\phi_{1}\rangle\dots |\phi_{N-1}\rangle,
\end{equation}
which is not Hermitian for $N\ge 3$. For separable states $\rho_{sep}=\rho_{1}\otimes\rho_{2}\otimes\dots\rho_{N}$ Eqn.~(\ref{2}) generalizes to
\begin{equation}\label{6}
    \mathrm{Tr}[\rho_{sep}  V_{N}]=\mathrm{Tr}[\rho_{a}\rho_{b}\dots\rho_{N}].
\end{equation}
For $N$ identical copies of a state, abbreviated as $\rho^{\otimes N}$, this becomes 
\begin{equation}\label{7}
    \mathrm{Tr}[\rho^{\otimes N}  V_{N}]=\mathrm{Tr}(\rho^{N})=\sum_{i}\lambda_{i}^{N}
\end{equation}
where $\lambda_{i}$ is the $i^{th}$ eigenvalue of $\rho$.  We note that for a set of $N$ identical pure states the expectation value is unity, while if the set is not homogeneous the expectation value will be less than unity.  This serves as a practical state discrimination test of multiple systems \footnote{See \cite{Jex04} for a related single-shot scheme.}.

For a $d-$dimensional system the spectrum of $\rho$ can be obtained from the ($d-1$) values of $\mathrm{Tr}(\rho^{N})$ for $N=2,3,\dots,d$ \cite{Horodecki:2002}.  Once the spectrum is known any nonlinear functional of the general form $\mathrm{Tr}[f(\rho)]$ can be computed through the corresponding function of the eigenvalues $\sum_{i}f(\lambda_{i})$. This follows from the fact that the trace is independent of the basis in which the density matrix is expressed. In the case when the system is entangled, knowledge of the spectrum of both the entangled state and the reduced subsystems can be used to test for entanglement through the majorization condition of Nielsen \cite{Nielsen:1999,Nielsen:2001}. 

In all of the above cases the desired measure was obtained from a single value: the expectation value of the swap operator. We now introduce a simple experiment to measure $\langle   V_{N}\rangle$ for a system of $N$ harmonic oscillators in an arbitrary state $\rho$.  

\begin{figure}[t] 
   \includegraphics[width=2in]{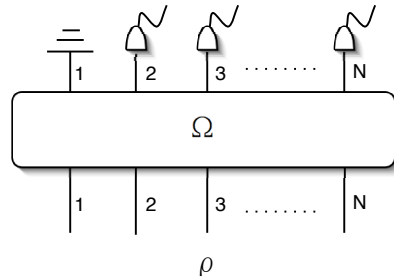} %figure figure
   \caption{Apparatus to measure $\mathrm{Tr}(\rho  V_{N})$. The evolution $ \Omega$ generates a discrete Fourier transformation of the $N$ oscillators.  Following that an energy measurement of $N-1$ oscillators is performed from which $\mathrm{Tr}(\rho  V_{N})$ is estimated.}
   \label{fig:1}
\end{figure}

The experiment, illustrated in Fig.~(\ref{fig:1}), 
is conducted in two stages. First the $N$ oscillators evolve under the action of the unitary operator $ \Omega$.  Following that a measurement is performed in the energy eigenbasis of each oscillator and the expectation value of an operator $  D$ is measured. The specific form of $ \Omega$ and $  D$ is such that
\begin{equation}\label{8}
    \mathrm{Tr}[ \Omega\,\rho\, \Omega^{\dag}  D]= \langle   V_{N}\rangle
\end{equation}
for all states $\rho$. Using the cyclic property of the trace this implies
\begin{equation}\label{9}
      V_{N}=  \Omega^{\dag}   D \Omega.
\end{equation}

Since, in general, $  V_{N}$ is not Hermitian, it is not strictly possible to associate $  D$ in Eqns~(\ref{8}) and (\ref{9}) with an observable. Nevertheless, it is possible to express a general operator as a \emph{weighted sum} of POVM elements as was done in \cite{Pregnell:2002a}. From the linearity of the trace, the expectation value of the operator corresponds to the same weighted sum of measured probabilities.  

%To illustrate, take, for example, the POVM element $\Pi_{k}=|\psi_{k}\rangle\langle \psi_{k}|$ with $|\psi_{k}\rangle=(|n\rangle+i^{k}|m\rangle)/\sqrt{2}$, where $|n\rangle$ forms an orthonormal basis. By a simple calculation it can be checked that the operator $|n\rangle\langle m|$ can be expressed as a linear combination of the POVM elements $(\Pi_{0}-\Pi_{2})/2+i(\Pi_{1}-\Pi_{3})/2$. As the operator $|n\rangle\langle m|$ form a basis it follows that any operator $ \Gamma=\sum_{nm}\gamma_{nm}|n\rangle\langle m|$ can be expressed as a weighted sum of POVM elements. By measuring the probability of the event associated with each POVM element in the expansion, the expectation value $\mathrm{Tr}(\rho \Gamma)$ of any operator can be measured. In general, not all POVM elements in the expansion belong to the same POVM which means that in general different experiments may have to be performed.  As we will show however, this is not the case here.  All the POVM elements in the expansion of $  D$ belong to the same POVM which means the settings of the experiment do not have to be changed from trial to trial.

From the action of $  V_{N}$ on the basis state in Eqn.~(\ref{5}) it can be shown that any operator $\mathcal O_{j}$ acting solely on the $j^{th}$ oscillator transforms as $  V_{N} \mathcal O_{j}   V_{N}^{\dag}\rightarrow \mathcal O_{j+1}$ (mod $N$).  Of particular importance are the creation and annihilation operators, $a^{\dag}_{j}$ and $a_{j}$ respectively, which satisfy the canonical commutation relation $[a_{i},a^{\dag}_{j}]=\delta_{i,j}$. Using the Kronecker delta function the transformation of the creation operators can be written as
\begin{equation}\label{10}
      V_{N} a^{\dag}_{j} V_{N}^{\dag}= \sum_{i}a^{\dag}_{i}\delta_{i,j+1}.
\end{equation}
Since the set of creation operators $\{a^{\dag}_{j}\}$ generates an orthonormal basis from the ground state $|0\rangle\otimes|0\rangle\otimes\dots\otimes|0\rangle$, Eqn.~(\ref{10}), along with the condition
\begin{equation}\label{10a}
    V_{N}|0\rangle\otimes|0\rangle\otimes\dots\otimes|0\rangle =|0\rangle\otimes|0\rangle\otimes\dots\otimes|0\rangle,
\end{equation}
is an equivalent definition of $V_{N}$. This is a simple but important result as it shows that the swap operation generates a \emph{linear} and \emph{unitary} transformation of the set of creation and annihilation operators. We now seek a solution to Eqn.~(\ref{9}) where the operators $\Omega$ and $D$ are of the \emph{same kind}. Specifically, we require
\begin{eqnarray}
      \Omega   a^{\dag}_{j}   \Omega^{\dag} &=& \sum_{i}a^{\dag}_{i}\Omega_{ij}\label{11}\\
      D   a^{\dag}_{j}   D^{\dag} &=& \sum_{i}a^{\dag}_{i}D_{ij},\label{12}
\end{eqnarray}
where the coefficients $\Omega_{ij}$ and $D_{ij}$, to be determined later, are elements of the unitary matrices $\mathbf\Omega$ and $\mathbf D$ respectively.

In general, any such unitary operator $U$ which transforms a set of $N$ creation operators linearly and unitarily 
\begin{equation}\label{10b}
      U a^{\dag}_{j}  U^{\dag}=\sum_{i} a_{i}^{\dag} U_{ij}
\end{equation}
is an element of the U($N$) Lie group \cite{Cornwell:Book}.  A property of the Lie group is that there corresponds a Hermitian operator $H$ of the general form
\begin{equation}\label{10c}
      H=\sum_{ij}\lambda_{ij}a^{\dag}_{i}a_{j},
\end{equation}
with $\lambda_{ij}=\lambda_{ji}^{*}$, which generates the group element $U=\exp(-iH)$. Noting that an arbitrary generator $H$ acting on the ground state is identically zero, we can Taylor expand $U$ as powers of $H$ and show that Eqn.~(\ref{10a}) is satisfied for any unitary $U$ of the Lie group, including the swap operator. 

The physical importance of Eqn~(\ref{11}) is that the evolution operator $\Omega$ can be implemented with a \emph{linear} coupling Hamiltonian of the form given in Eqn.~(\ref{10c}). \citeauthor{Reck:1994}  have shown how any such multi-particle coupling Hamiltonian can be implemented using a sequence of two particle interactions \cite{Reck:1994}. In an optics context this corresponds to an array of beam-splitters and phase-shifters.  As for the measurement, our requirement that $D$ be an element of U($N$) as-well-as correspond to a measurement in the energy basis implies that it must be generated by an operator of the form $H_{D}=\sum_{j}\theta_{j} a^{\dag}_{j}a_{j}$.  Explicitly, we require 
\begin{eqnarray}\label{14}
      D=&\exp[-i\theta_{1}  a^{\dag}_{1}  a_{1}]\otimes\exp[-i\theta_{2}  a^{\dag}_{2}  a_{2}]\otimes\dots\nonumber\\
   & \dots\otimes\exp[-i\theta_{N}  a^{\dag}_{N}  a_{N}].
\end{eqnarray}
where $\theta_{j}$ are free parameters yet to be determined. 

To derive the specific form of the matrix elements $\Omega_{ij}$ and the coefficients $\theta_{j}$ we substitute Eqn.~(\ref{9}) into (\ref{10}) and derive, with the help of (\ref{11}) and (\ref{12}), the \emph{matrix} equation 
\begin{equation}\label{13}
    \mathbf V_{N}=\mathbf\Omega^{\dag}\mathbf D\mathbf\Omega
\end{equation}
where $[\mathbf V_{N}]_{ij}:=\delta_{i,j+1}$.  We note that the commutation relation between the creation and annihilation operators implies
\begin{equation}\label{x}
    \exp(-i\phi \,a^{\dag}_{j}a_{j})\,a^{\dag}_{k}\exp(i\phi \,a^{\dag}_{j}a_{j})=a^{\dag}_{k}\exp(-i\phi\,\delta_{j,k}),
\end{equation}
from which it is straightforward to show from (\ref{14}) and (\ref{12}) that the matrix $\mathbf D$ is diagonal with elements $D_{jj}=\exp(-i\theta_{j})$.  The matrix $\mathbf\Omega$ in (\ref{13}) then is such that it diagonalises $\mathbf V_{\!N}$.  Using standard techniques we can solve Eqn.~(\ref{13}) to give
\begin{eqnarray}
    \Omega_{ij}&=&\omega^{ij}/\sqrt{N}\label{15}\\
    D_{jj}&=&\omega^{j-1}\label{16}
\end{eqnarray}
where $\omega=\exp(i2\pi/N)$ is the $N^{th}$ root of unity.  We see that the matrix $\mathbf \Omega$ is a discrete Fourier transformation and defines the action of the unitary operator $ \Omega$.  The unknown phases in (\ref{14}) are found from (\ref{16}) to be 
\begin{equation}\label{17}
    \theta_{j}=2\pi (j-1)/N,\quad\quad j=1,2,\dots,N
\end{equation}
and characterize the operator $  D$ in (\ref{14}).  To express this in terms of POVM elements we rewrite the energy operator $a^{\dag}_{j}a_{j}$ in the energy basis of the $j^{th}$ oscillator as $\sum_{n_{j}}n_{j}|n_{j}\rangle\langle n_{j}|$. With this (\ref{14}) becomes
\begin{equation}\label{14a}
      D=\sum_{\{n_{i}\}} w_{1}(n_{1})\dots w_{N}(n_{N})|n_{1}\dots n_{N}\rangle\langle n_{1}\dots n_{N}|
\end{equation}
where 
\begin{equation}\label{14aa}
    w_{j}(n_{j})=\exp(-i\theta_{j}n_{j})
\end{equation}
is a complex weighting coefficient. In the context of a measurement the projector $|n_{1}\dots n_{N}\rangle\langle n_{1}\dots n_{N}|$ is associated with the joint measurement outcome $(n_{1},n_{2},\dots,n_{N})$, which is interpreted as the outcome $n_{1}$ occurring at the first oscillator, $n_{2}$ at the second oscillator and so on. The probability of observing the event $(n_{1},n_{2},\dots,n_{N})$ is given by the overlap of the state with the associated projector.  For the state $ \Omega\rho \Omega^{\dag}$ the joint probability is
\begin{equation}\label{14b}
    \mathrm{Pr}(n_{1},n_{2},\dots,n_{N})=\mathrm{Tr}( \Omega\rho \Omega^{\dag}|n_{1}\dots n_{N}\rangle\langle n_{1}\dots n_{N}|).
\end{equation}
From (\ref{14a}) and (\ref{14b}) the expectation value $\mathrm{Tr}( \Omega\rho \Omega^{\dag}  D)$ can be expressed as a (complex) weighted sum of measured probabilities, 
\begin{eqnarray}\label{14c}
    \lefteqn{\mathrm{Tr}( \Omega\rho \Omega^{\dag}  D)}\\
    &=&\sum_{\{n_{i}\}} w_{1}(n_{1})w_{2}(n_{2})\dots w_{N}(n_{N})\mathrm{Pr}(n_{1},n_{2},\dots,n_{N})\nonumber
\end{eqnarray}
which is, by definition, the expectation value $\langle  V_{N}\rangle$. Interestingly, $\theta_{1}=0$ which means that the weighting function $w_{1}(n_{1})=1$ and is independent of the measurement outcome $n_{1}$ of oscillator one. Accordingly, no information about the state of oscillator one (after the interaction) is used in the estimate of $\langle  V_{N}\rangle$.  To simplify matters experimentally, no measurement need be performed on this oscillator. 

\begin{figure}[t] %  figure placement: here, top, bottom, or page
   \includegraphics[width=2in]{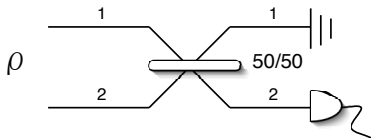} 
   \caption{Apparatus to measure $\mathrm{Tr}(\rho  V_{2} )$. No measurement is performed on oscillator one, while detector two measures the average value of the parity. The two oscillators interact via a 50/50 coupling device.} 
   \label{fig:2}
\end{figure}

To illustrate the practicality of this apparatus we will briefly discuss the simplest case which is when $N=2$.  The circuit diagram of this experiment is illustrated in Fig.~\ref{fig:2}.
As previously mentioned, for separable states $\rho_{1}$ and $\rho_{2}$ the experiment estimates the overlap $\mathrm{Tr}(\rho_{1}\rho_{2} )$ between two unknown states.  By changing the states at the input of the apparatus we can choose the quantity we wish to measure, ranging from the purity when both systems are in identical states $\rho_{i}$, through to the fidelity $\langle \alpha|\rho|\alpha\rangle$ if one of the states is pure $|\alpha\rangle$.  On the other hand, for entangled states the experiment acts like an entanglement witness giving a negative value for some entangled states while ensuring a positive value for all separable states.

From Eqns~(\ref{11}) and (\ref{15}) the required unitary transformation is of the form 
\begin{equation}\label{18}
     \Omega\left(\begin{array}{c}a_{1}^{\dag} \\a_{2}^{\dag}\end{array}\right) \Omega^{\dag}= \frac{1}{\sqrt{2}}\left(\begin{array}{cc}1 & 1 \\1 & -1\end{array}\right) \left(\begin{array}{c}a_{1}^{\dag} \\a_{2}^{\dag}\end{array}\right).
\end{equation}
This specific unitary is generated by an interaction Hamiltonian of the form $i\kappa(a^{\dag}_{1}a_{2}-a_{1}a_{2}^{\dag})$ applied for a time $\pi/(4\kappa)$, where $\kappa$ is the interaction strength. In optics, where the two oscillators are realized by different spatial modes of the quantized field for example, this can be achieved with a simple 50-50 beam-splitter. In the case where the oscillators are distinguished by different polarization and/or temporal modes, then additional polarizing beam-splitters and/or time delays would be required. Following the interaction a measurement is performed in the energy basis of each oscillator and the probability $\mathrm{Pr}(n_{1},n_{2})$ is observed after repeated trials. To extract $\langle V_{2}\rangle$ the distribution is weighted against the coefficients $w_{1}(n_{1})=1$ and $w_{2}(n_{2})=(-1)^{n_{2}}$ given by (\ref{14aa}) and (\ref{17}). The result is
\begin{equation}\label{19}
    \langle V_{2}\rangle =\sum_{n_{2}}(-1)^{n_{2}}\mathrm{Pr}(n_{2})
\end{equation}
where $\mathrm{Pr}(n_{2})=\sum_{n_{1}}\mathrm{Pr}(n_{1},n_{2})$, and is independent of $n_{1}$, the outcome of the measurement on oscillator one. To simplify matters experimentally only the distribution $\mathrm{Pr}_{+}=\sum_{n_{2}} \mathrm{Pr}(n_{2})$ and $\mathrm{Pr}_{-}=1- \mathrm{Pr}_{+}$ of the second oscillator need be measured. The expectation value  $\langle V_{2}\rangle$ can then be obtained from the difference $\mathrm{Pr}_{+}-\mathrm{Pr}_{-}$, which is known as the average parity and corresponds to the zero of the Wigner function\cite{Banaszek:1999}. Similar simplifications reside in the measurement of $\langle V_{N}\rangle$ for higher values of $N$. It is a surprising result that such a range of meaningful measures can be obtained from an experiment that requires no non-linearity, no interferometers, just a linear interaction and a single parity measurement.  

We note that higher order moments of the swap operator $\langle(V_{N})^{k}\rangle$ can also be obtained from the general apparatus illustrated in Fig~\ref{fig:1}.  This can be seen by writing $(V_{N})^{k}= \Omega^{\dag}  D^{k} \Omega$.  Physically, this corresponds to the same evolution $ \Omega$ of the state $\rho$ followed by a measurement of $D^{k}$ which can be written as 
\begin{eqnarray}\label{y}
      D^{k}=&\exp[-ik\theta_{1}  a^{\dag}_{1}  a_{1}]\otimes\exp[-ik\theta_{2}  a^{\dag}_{2}  a_{2}]\otimes\dots\nonumber\\
   & \dots\otimes\exp[-ik\theta_{N}  a^{\dag}_{N}  a_{N}].
\end{eqnarray}
Repeating the calculation it is seen that this corresponds to measuring the same probability distribution $\mathrm{Pr}(n_{1},n_{2}\dots n_{N})$, however now the distribution is weighting by the functions $w_{j}(n_{j},k)=[w_{j}(n_{j})]^{k}$.  

The measurement procedure introduced here generalizes to measure the expectation value $\mathrm{Tr}(\rho U )$ of \emph{any} unitary operator $U$ that transforms the creation operators of $N$ harmonic oscillators linearly. The key is that the associated unitary matrix $\mathbf U$ can always be diagonalized as $\mathbf\Omega^{\dag}\mathbf D\mathbf\Omega$ where both $\mathbf\Omega$ and $\mathbf D$ are associated with the unitary evolution $ \Omega$ and the measurement $D$ through Eqns~ (\ref{11}) and (\ref{12}) respectively. The expectation value $\mathrm{Tr}(\rho U )$ is then given by the weighted sum of probabilities as in (\ref{14c}), where the specific values of the phases $\theta_{j}$ are determined from the diagonal matrix $\mathbf D$.

In conclusion, we have introduced a procedure to directly measure the quantity $\mathrm{Tr}(\rho U)$ for any unitary operator generated by a linear coupling Hamiltonian. The procedure removes the experimentally challenging step of entangling an ancilla to the $N$ systems which was present in previous proposals.  The result is a more practical procedure involving only linear interactions between $N$ oscillators and local energy measurements. In the case where the unitary operator is the $N$-particle swap operator, the measurement corresponds to a range of useful measures depending on the input state, including the purity, the fidelity, the overlap, an entanglement witness and generalized non-linear functionals.

\bibliography{General}    
\end{document}